# An Adaptive Proton FLASH Therapy Using Modularized Pin Ridge Filter

Ahmal Jawad Zafar[1], Xiaofeng Yang[1], Zachary Diamond[1], Tian Sibo[1], David Yu[1], Pretesh R. Patel[1] and Jun Zhou[1]*

[1]Department of Radiation Oncology and Winship Cancer Institute,
Emory University, Atlanta, GA, 30322, USA
*Email: jun.zhou@emory.edu

## Abstract:

In this paper, we proposed a method to optimize adaptive proton FLASH therapy (ADP-FLASH) using modularized pin-ridge filters (pRFs) by recycling module pins from the initial plan while reducing pRF adjustments in adaptive FLASH planning. Initially, single energy (250 MeV) FLASH-pRF plans were created using pencil beam directions (PBDs) from initial IMPT plans on the planning CT (pCT). PBDs are classified as new/changed (ΔE > 5 MeV) or unchanged by comparing spot maps for targets between pCT and re-CT. We used an iterative least-square regression model to identify recyclable PBDs with minimal relative changes to spot MU weighting. Two PBDs with the least square error were retrieved per iteration and added to the background plan, and the remaining PBDs were reoptimized for the adaptive plan in subsequent iterations. The method was validated on three liver SBRT cases (50 Gy in 5 fractions) by comparing various dosimetric parameters across initial pRF plans on pCT, re-CT and the ADP-FLASH-pRF plans on re-CT. $V_{100}$ for initial-pRF plans on pCT, re-CT, and ADP-FLASH-pRF plans for the three cases were as follows: (93.7%, 89.2%, 91.4%), (93.5%, 60.2%, 91.7%), (97.3%, 69.9%, 98.8%). We observe a decline in plan quality when applying the initial pRF to the re-CT, whereas the ADP-FLASH-pRF approach restores quality comparable to the initial pRF on the pCT. FLASH effect of the initial pRF and ADP pRF plans were evaluated with a dose and dose rate threshold of 1Gy and 40Gy/s, respectively, using the FLASH effectiveness model. The proposed method recycled 91.2%, 71%, and 64.7% of PBDs from initial pRF plans for the three cases while maintaining all clinical goals and preserving FLASH effects across all cases.

# 1. Introduction:

Radiotherapy has been a pivotal part of cancer treatment for many decades. However, from the beginning the use of high-energy radiation to target cancerous cells, while avoiding damage to healthy tissues has been one of the leading challenges in the field of radiotherapy. FLASH radiotherapy (FLASH-RT), has shown promising results at an ultrahigh dose rate, generally greater than 40 Gy/s. Unlike conventional low dose-rate therapies, FLASH-RT has demonstrated a reduction in damage to healthy tissue without compromising the effectiveness of tumor treatment[1-4]. An animal study of FLASH-RT utilizing a high enough dose rate has shown promising results in effectively sparing various organs at risk (OARs), such as abdominal tissues, lungs, brain, and skin[1,2,5-7]. Additionally, multiple clinical trials on human patients have also shown the validation of FLASH-RT. The first clinical trial for a patient with T-cell cutaneous lymphoma on the effects of FLASH by treating with electron beams has demonstrated promising outcomes for both normal skin and the tumor[8]. Similarly, the FAST-01 trail has also been proven to provide further evidence of the feasibility and efficacy of proton FLASH-RT for patients with multiple bone metastases[9,10]. Ongoing research into the effects of proton and electron FLASH-RT on bone metastases in the chest[11] and skin melanoma metastases[12] is expanding our understanding and enhancing the potential applications of FLASH-RT.

Proton beams, known for their narrow Bragg peaks (BPs), are preferred in FLASH-RT because they offer precise targeting of tumors and superior sparing of surrounding healthy tissues[13]. Furthermore, a delivery method based on pencil beam Direction (PBD) for Intensity-Modulated Proton Therapy (IMPT) is frequently utilized in proton therapy. It enhances the precision of targeting complex tumor shapes while minimizing radiation exposure to OARs, which significantly improves the safety and effectiveness of treatment[14]. However, in a clinical cyclotron, achieving the spread-out Bragg peaks (SOBPs) necessary for precision in IMPT typically involves modulating proton beam energies using an energy degrader and energy selection system. This modulation can considerably reduce the beam current, hindering the delivery of proton beams at ultra-high FLASH dose rates[15]. To avoid this inefficiency, there is a push to use high-energy proton transmission beams (TBs), allowing the coverage of the target area while keeping the high-energy points (BPs) outside the patient, avoiding the need for inefficient energy modulators. While this approach has shown promising results in achieving the desired FLASH effects[16-19], it has also posed risks, such as potential overexposure of normal tissues near the distal edge of the target to high exit doses. One approach to improve the efficiency of this method is combining high-energy proton TBs with BPs from conventional dose rates. It aims to provide comprehensive dose coverage at the target edges while sparing the surrounding normal tissues[20]. Another development includes using step-shaped bar ridge filters to spread out the single-energy SOBP, enhancing OARs sparing, and achieving more homogenous target dosing for FLASH planning[21].

To improve the effectiveness of BPs in FLASH-RT, pin-shaped ridge filters (pRFs) are engineered with ridge pins having step-like shapes aimed at generating a uniform SOBP for each PBD used in IMPT[22-25]. Additionally, retracting the ranges of the highest energy proton beams while adjusting proton ranges to match the distal edge of the target can eliminate the exit dose, enhancing the protection of OARs while maintaining FLASH dose rate delivery[25]. Our previous studies have demonstrated a method based on the adjustable modulation of the dose distribution[26]. Like traditional PBD-IMPT methods, our method provides flexibility in the optimization of an IMPT downstream (IMPT-DS) plan followed by direct conversion to a corresponding FLASH plan, which has led to a more streamlined pRF design featuring coarser pin resolution, simplifying production and enabling reusable modules. Utilizing a strategy for pRF design leverages PBD-based spot reduction to create a multi-energy IMPT plan. This plan uses a downstream energy modulation approach (IMPT-DS) which is effectively transformed into streamlined pRFs. This enables us to incorporate our FLASH planning technique into existing clinical treatment planning systems (TPS).

Conventional proton therapy shows high susceptibility to anatomical changes, whereas BPs in FLASH-RT demonstrate even greater sensitivity to such uncertainties. The pre-printed/built pRFs are less flexible in adapting to these changes, resulting in inferior dosimetry in ultra-hypofractionated FLASH treatments. IMPT-pRF planning can be complex and time-consuming[27] and the pRF printing is cumbersome. This study introduces a novel IMPT-pRF planning method that leverages previously designed pRF, aiming to enhance planning efficiency while maintaining high plan quality. The methodology involves creating an initial IMPT plan and corresponding pRF plan based on the patient's initial planning CT (pCT). Upon acquiring a new CT (re-CT) that reflects anatomical changes, adaptive (ADP) pRF is created while minimizing changes in the original RF pins by recycling the pins from the pRF plan on pCT[28]. This approach improves the time and cost of replanning by allowing a significant number of pins to be reused from the initial pRF plan for the ADP-pRF plan.

This paper is organized into five sections. Section 2 outlines the method used in this study, with Section 2.1 detailing the Flash planning framework for designing an efficient pRF configuration, and the modularization scheme for pRF, focusing on the parameters critical for achieving optimal Flash planning outcomes. Section 2.2 describes the process for selecting specific PBDs that remain constant during adaptations of the Flash plan in response to anatomical changes. Section 3 presents the results obtained from these proposed methods, while Section 4 offers a discussion based on these results. Finally, Section 5 summarizes and concludes the findings.

## 2. Methods and Materials:

This section presents the comprehensive methodology adopted for this study, including the development of the Flash planning framework and the modularization of the pin-RF configuration. We detail the approach used to ensure effective FLASH planning and the specific parameters involved in optimizing the design. Additionally, we describe the most crucial process of this study for selecting persistent PBDs during anatomical adaptations in the context of stereotactic ablative body radiotherapy (SABR) liver study. The subsequent subsections break down these components in detail.

## 2.1 pRF Modularization for FLASH Planning:

We have previously published work involving transforming pRFs from an IMPT plan[26]. The modulation of the depth dose curve for a single energy proton beam in the IMPT plan, along with the same PBD, is based on utilizing step thickness and surface areas of each ridge pin. For each PBD in the IMPT plan, the BP depths $R_i$ and $R_{i-1}$ determines the water equivalent thickness (WETs), $T_i$, of step $i$ (where $i > 1$) in a ridge pin as described in equation (1).

$$T_i = R_i - R_{i-1} \tag{1}$$

WETs are further adjusted by introducing factors, $T_{RS}$, WET thickness pin-RF base, and $T_B$, WETs of Range shifter (RS), for each field, equation (2).

$$T_1 = R_0 - T_{RS} - T_B \tag{2}$$

$$\bar{w}_i = \frac{w_i}{\sum_{i=1}^{N} w_i} \tag{3}$$

$$\bar{w}_i = \frac{L_{i-1}^2 - L_i^2}{L_1^2} \tag{4}$$

To design streamlined pin-RFs with coarse resolution and sparse pin distribution, an inverse planning framework within RayStation 10B for FLASH planning was established[26]. This framework starts with creating an IMPT plan with multiple energy layers (IMPT-DS plan). It then removes low-weighted energy layers and PBDs, translating the IMPT-DS plan into pRFs with single-energy beam configurations for the FLASH plan (IMPT-pRF).

For FLASH planning, the pRFs are designed using streamlined pyramid-shaped ridge pins. These pyramid-shaped pins are assembled using cuboid-shaped unit modules chosen from a set of reusable modules, each with a specific width and WET. In the RF, each pin's shape is transformed via the IMPT-DS plan during the FLASH planning process. The settings of the IMPT-DS plan regulate the pin spatial resolution and sparsity of ridge pins, thereby determining the shapes and quantities of the unit modules necessary for assembly.

The 1 mm step width resolution of the ridge pin aligned perfectly with the FLASH planning's streamlined pRF previously published. Achieving this coarse resolution requires a strategic reduction in the number of steps per ridge pin, accomplished through the nested PBD-based spot reduction process in IMPT planning. This process is crucial as the step widths of a ridge pin are determined from the spot weights along the corresponding PBD of the IMPT-DS plan, as described in equations (3) and (4). Consequently, the calculated step widths are rounded to the nearest whole number in millimeters. A detailed explanation of this resolution can be found in a previous study[26]. Furthermore, the 1 mm resolution aligns with the minimum dose grid resolution used by most commercial TPS and corresponds to the optimized pRF resolution adopted in our prior implementation of the streamlined FLASH planning framework.

In addition, the factors of WET and the width of each step determined the overall resolution of the pyramid shaped in pRF. The WET of each step reflects the variation in BP depths between two successive energies along the PBD, which matches the energy layer spacing used in the IMPT-DS planning process. This study established a 5 mm energy layer spacing in the IMPT-DS plan, ensuring the WET of the unit modules, also featuring this resolution.

In this study, the unit modules are designed with six specific widths, starting from 1 mm to 6 mm in 1 mm increments, with the bottom step width, $L_1$, fixed at 6 mm. Each module in the set features a WET of 5 mm and discrete widths from 1 to 6 mm, as shown in Figure 1. This modular design enables the construction of any pRF step by stacking unit modules to match the desired step width.

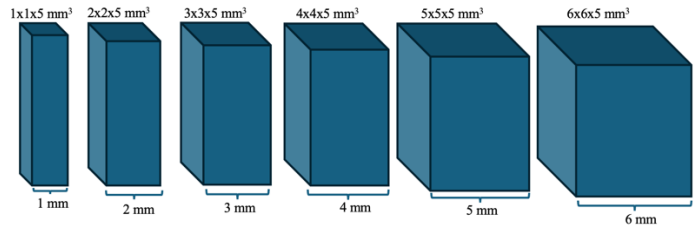

Figure 1. Unit modules of six sizes

Moreover, to assemble the pRFs, reducing the total number of ridge pins is essential. This is achieved in the IMPT-DS plan by iterative spot reduction of low-weighted PBDs with the total sum of spot monitor units (MUs) less than a certain predefined threshold. It is crucial to recognize that in the translated IMPT-pRF plan, the spot MU for PBD represents the sum of spot MUs for that PBD in the IMPT plan. Therefore, to achieve the FLASH effect, the MU threshold needs to be high enough so that the minimum spot weight of the IMPT-pRF can be delivered at a high beam current. In this work, the MU threshold was set to 300 MU, guaranteeing the deliverability of the IMPT-RF at a beam current of 500 nA and 0.5 ms of a minimum spot duration time.

## 2.2 Adaptation of PBD:

As proton therapy is sensitive to anatomy changes, a FLASH RF plan developed based on initial pCT may need adaptation to new anatomy. The strategy adopted here is to reoptimize the plan while minimizing the RF changes (recycle the existing RF pins) for the new anatomy. This section outlines a dynamic recycle technique that iteratively expands the reusable pins to minimize the RF pin change. The whole process is described as:

a) **Initializing Spot map:** The initialized spots on pCT and re-CT are mapped to identify which of the PBDs among the plans are 'unchanged' in terms of maximum and minimum spot energy. Identifying the new and changed PBDs caused by anatomical changes, as opposed to those that remain 'unchanged' on the reCT. A threshold for change is set at a 5 MeV difference in either the maximum or minimum spot energy within each PBD, as represented by:

$$PBD\ tag = \begin{cases} \text{"changed"} & if\ \Delta E_{distal} \geq 5MeV \\ \text{"changed"} & if\ \Delta E_{proximal} \geq 5MeV \\ \text{"unchanged"} & otherwise \end{cases}$$

The unchanged PBDs are identified and marked in Figure 2(a) with red open rectangles and enclosed by a red dotted box. Any PBDs that do not fall into the criteria of the above equation are termed 'new/changed' PBDs (pink rectangles).

b) **IMPT-DS plan on pCT**: Range shifter (RS) plates of various thicknesses are placed downstream of the nozzle to shift the Bragg peak depth of the proton beam in an IMPT-DS plan[25,29,30]. Similar to the approach taken in conventional IMPT planning, energy level selections and spot weights for IMPT-DS planning also result from a constrained optimization process. This method ensures the target receives the necessary dose while reducing exposure to surrounding healthy tissues. After constructing a fully optimized IMPT-DS plan using the RayStation planning platform on pCT, we compare its PBDs with previously processed PBDs (step (a)) to identify the spatially common PBDs. This plan serves as the foundation for extracting PBDs that will be marked as a set of 'Recycled PBDs' (Background in Fig. 2($c_3$)).

c) **Selection Process:** Initially, all PBDs, new/changed and unchanged, are placed in an adaptive beamset (Fig. 2($c_1$)). An iterative process using the optimization of the least squares objective function calculates the error of all PBDs and selects the two PBDs with the minimum error in each iteration, placing them in the background (Fig. 2(c$_3$)). The mathematical expression for the error function is:

$$min_\beta \|Y\beta - X\|^2 \tag{5}$$

Here, β represents the scaling factor for the least squares fit. *X* and *Y* denote the spot ion MUs normalized by the total weight of all spots within the corresponding PBD, as shown in Figures 2(b) and 2(c$_1$) dotted red box, respectively. For any i$^{th}$ spot in the j$^{th}$ PBD, the expressions for the weighted ion MUs are expressed as:

$$y_{ij} = \frac{(ion\_MU)_i}{\sum_j (ion\_MU)_j} \quad , \quad y_{ij} \in Y$$

The PBDs from the Adaptive DS plan (Figure 2(c$_1$)) with a minimum error are prioritized for transfer in the Background (Figure 2(c$_3$)). The rest of the PBDs are categorized as 'Remaining' in Figure 2(c$_2$). The 'Remaining' PBDs are optimized and become the adaptive DS plan (Figure 2(c$_1$)) for the next iteration. This iterative approach continues until sufficient selected PBDs are recycled without compromising the quality of dosimetry and clinical objectives of the plan. This approach ensures that the structural integrity of each recycled PBD, determined by the proportion of its spot weighting, remains unchanged in the final plan. Throughout the iterative optimization, the PBDs are recycled, and the quality of the adaptive plan is maintained.

In summary, the process unfolds in three key steps. First, we aimed to identify and mark the PBDs that are impacted by these anatomical changes. In the second step, an initial IMPT-DS plan was constructed from the pCT, representing the original plan before any anatomical changes occurred. Finally, in the third step, we used a least squares linear regression model to develop an ADP DS plan based on PBDs from the original DS plan, aiming for minimum changes as it will eventually result in an ADP pRF plan, which was built using the recycled PBDs from the plan designed in the first step, allowing us to apply it to the re-CT.

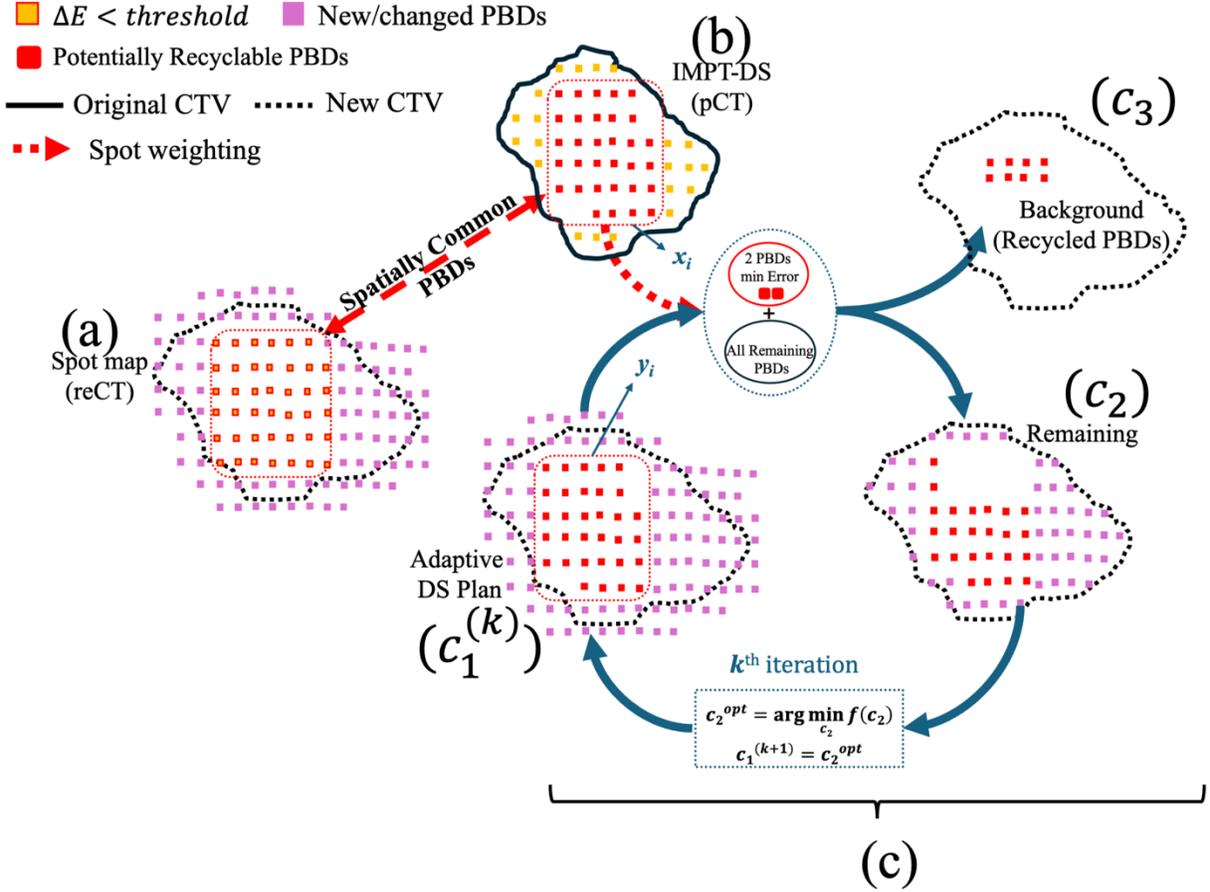

**Figure 2**. A schematic diagram of the beam's eye view of the CTV, detailing the selection process for iteratively recycling PBDs. Spot maps from the pCT and re-CT were compared to identify and mark the set of new/changed PBDs, and unchanged PBDs based on the threshold ΔE > 5 MeV. Both new/changed and unchanged PBDs were initially placed in the **Adaptive DS** plan. During iterative recycling, a set of two PBDs per iteration are retrieved by least squares regression to minimize changes in pin structure and placed in the background **($c_3$)** plan while the Remaining **($c_2$)** PBDs are reoptimized and become **Adaptive DS** plan from next iteration.

## 2.3 Flash Effect:

In this study, we have used the effectiveness model to quantify the FLASH dose of the IMPT-pRF plan[31]. This model hypothesizes that the FLASH effect is initiated in a normal tissue voxel if the dose ($\Delta d$) delivered within a specific time window ($\Delta t$) is more significant than a dose threshold ($D_0$), $\Delta d \geq D_0$ is administered at an average dose rate that exceeds a dose rate threshold, $\Delta d/\Delta t \geq \dot{D}_0$. The FLASH effect remains active throughout the triggering window and an additional persistence period. The dose delivered with the FLASH effect is considered biologically safe, providing less toxicity by applying an effective factor (less than 1) [31].

In this investigation, the parameters for dose rate ($\dot{D}_0$), persistence time, and effective factor were set to 40 Gy/s, 200 ms, and 0.67, respectively, based on previously reported values[32]. The effective factor of 0.67 implies a 33% reduction in the dose delivered to normal tissues when the FLASH effect is active. Given the dependency of the FLASH effect on the dose threshold $D_0$, values within the range of 1–25 Gy can be used to evaluate the effect. However, $D_0$ must not exceed the average beam dose per fraction. Furthermore, the Varian research group demonstrated the scanning pattern with a scanning speed of 10 mm/ms commissioned to assess the FLASH effective dose[19].

The clinical benefits were studied by comparing the FLASH dose of IMPT-pRF to the dose of a conventional IMPT (IMPT-CONV) plan in target and OAR tissues and quantifying the resulting dose reduction. As FLASH presumably does not affect tumor cells, no dose reduction is applied to the GTV region. Nevertheless, in regions other than the GTV, the planning target volume (PTV), referred to as PTV-GTV, is particularly interesting. This region, which may contain both healthy tissue and tumor cells and is exposed to high doses of radiation, could exhibit a clinically significant differential response if the FLASH effect operates at the cellular level. Therefore, we evaluated the clinical benefits of FLASH in the PTV-GTV region and its impact on other surrounding normal tissues.

## 2.4 Patient Study:

The proposed adaptive method for the modularized pRF was applied to three liver cancer cases that had undergone adaptation in conventional proton SBRT. For cases 1, 2, and 3, the clinical target volume (CTV) on the pCT measured 48.4 cc, 25.2 cc, and 47.2 cc, respectively. Additionally, the CTVs on the re-CTs, which differ in shape and position from those CTVs on pCTs, measured 48.9 cc, 25.9 cc, and 46.2 cc for cases 1, 2, and 3, respectively. In accordance with the SABR protocol[33], the prescription doses were set to 50 Gy and delivered in 5 fractions. To achieve precise dose conformity for the target while protecting normal tissues, the beam angles in the FLASH planning were taken from the original IMPT-CONV plans. Two beams were used for cases 1 and 2, while three beams were used for case 3. The treatment goal was to ensure that 95% of the CTV received the total prescription dose (V100 ≥ 95%) required for the CTV. Additionally, the maximum dose (Dmax) to the CTV was limited to 140% of the prescribed dose, i.e., (Dmax ≤ 70 Gy). The quality of the ADP-pRF plan was assessed by comparing it to the corresponding initial pRF plan, centering on key dose indicators and target dose conformity.

We use the FLASH effectiveness model, utilizing the parameters mentioned in section 2.3, to measure FLASH effects and dose distribution. The clinical benefits of the FLASH effect, resulting in the proposed method of ADP IMPT-pRF plan, were assessed by comparing the concerned dose indicators of the FLASH dose distribution with conventional IMPT plans. Finally, the same beam arrangements, prescription dose, and objective targets were adopted from IMPT plans on pCT and re-CT to ensure consistency with all treatment plans.

# 3. Results

## 3.1 Adaptation Process:

The total number of ridge pins in the initial pRF plan on pCT for cases 1, 2, and 3 were 52, 69, and 73, respectively. For each case, the initial pRF plan, designed on pCT, served as a benchmark for comparison with the ADP-pRF plan. The analysis of this recycling PBD method was conducted by comparing three scenarios. First, the initial pRF plan was developed based on the pCT. Second, the initial pRF plan was applied to the re-CT to evaluate plan quality and observe differences. Finally, an ADP-pRF plan was constructed on the re-CT and compared with the initial pRF plan. Across all three plans, the results demonstrate the necessity of developing a new adaptation process based on the proposed method.

### *3.1.1 Initial pRF designed for pCT on re-CT:*

In each case, the initial pRF plan, translated from the IMPT-DS plan, met all dosimetric criteria, ensuring that 98% of the CTV received the full prescription dose. Re-evaluation of the initial pRF plan on re-CT revealed significant discrepancies in dose coverage (V100). Across three evaluated cases, the V100, representing the target volume receiving 100% of the prescribed dose, demonstrated consistent decline when evaluated on re-CT compared to initial pCT assessments. In case 1, the initial pRF on pCT achieved a V100 of 93.7%, which decreased to 89.2% when recalculated on re-CT. Case 2 exhibited a more pronounced reduction, with V100 values dropping from 93.5% (pCT) to 60.2% (re-CT). Similarly, Case 3 showed a marked decline, transitioning from 97.3% on pCT to 69.9% on re-CT. These trends suggest anatomical or positional changes between scans compromise target coverage over time. The Homogeneity Index (HI), a quantification of dose uniformity within the target volume, was also analyzed. Higher HI values indicate greater dose heterogeneity, which correlated with the observed reductions in V100.

$$HI = \frac{D_{2\%} + D_{98\%}}{D_{50\%}}$$

The most significant increase was observed in case 3, where the HI rose to 0.8, indicating a markedly less uniform dose distribution when applying the initial pRF to re-CT. Figure 3(a) and (c) highlight the quality differences in the treatment plan for case 1 by comparing the two-dimensional dose distributions on the initial pRF plan generated using the pCT and the re-CT. A similar trend is evident in the dose-volume histogram (DVH) shown in Figure 3(b), where the coverage of the clinical target volume (CTV) is significantly lower in the re-CT compared to the original plan based on the pCT.

### *3.1.2 ADP pRF designed on re-CT:*

The suboptimal dosimetric performance observed in the initial pRF plan on re-CT highlighted the necessity of implementing the proposed method, which iteratively recycled PBDs from the initial pRF plan to design ADP-pRF plans on re-CT. On average, 75% of ridge pins were recycled from the initial pRF plan during the design of the ADP-pRFs. Case-specific analysis revealed that the highest recycling rate occurred in case 1, with 91.2% of pins being reused. In contrast, case 2 achieved a recycling of 64.7%, and case 3 achieved 71%. A comparable level of dose conformity was achieved in the ADP-pRF plan compared to the initial pRF plan while recycling a substantial portion of original pins from the initial pRF plan, as demonstrated in Figure 1(d), for case 1. In the ADP-pRF plan on re-CT, the V100 values of the CTV for case 1, case 2, and case 3 were 91.40%, 91.7%, and 98.8%, respectively. The difference in V100 value between the initial pRF plan and our approach was no more than 2.3%, as observed in case 1. In contrast, case 3 demonstrated a slight improvement over the initial value, indicating enhanced CTV coverage. The ADP-pRF plan maintained dose distribution uniformity across all cases, Figure 1(d). As detailed in Table 1, our approach achieved comparable performance to the initial pRF plan on pCT compared to the initial pRF plan recalculated on re-CT. Specifically, Table 1 quantifies key metrics for the three scenarios: (1) initial pRF on pCT, (2) initial pRF on re-CT, and (3) ADP-pRF on re-CT. For OAR sparing, the variations in dose metrics, V0.5cc to the stomach across all cases were generally minimal. Case 3 demonstrated improvement, with a 4.8 Gy reduction compared to the initial pRF plan. A notable exception was observed in case 3 where the Deudenum (V0.5cc) increased by 2.03 Gy, marginally exceeding the BR001 protocol tolerance for the ADP-pRF plan. This increase can be attributed to the Deudenum being situated in a high-dose region in re-CT, making it more vulnerable to additional dose spillage. The D700cc for the liver (excluding the gross tumor volume, liver-GTV), defined as the dose delivered to the liver-GTV volume minus 700 cc, showed minimal variation (≤ 0.5 Gy) between the initial pRF and ADP-pRF plans across all cases. For the esophagus (V0.03cc), doses increased slightly in all cases, although, the values remained within protocol tolerance limits.

Table 1. Dosimetric parameters observed for OARs for all three case studies. Each case is divided into three plans, starting from the initial plan of pCT, followed by re-CT, and lastly, the ADP-pRF plan constructed by using the iterative PBD recycling method.

|  | case 1 | | | case 2 | | | case 3 | | |
|---|---|---|---|---|---|---|---|---|---|
|  | Initial pRF (pCT) | Initial pRF (re-CT) | ADP-pRF (re-CT) | Initial pRF (pCT) | Initial pRF (re-CT) | ADP-pRF (re-CT) | Initial pRF (pCT) | Initial pRF (re-CT) | ADP-pRF (re-CT) |
| $V_{100}$ | 93.7% | 89.2% | 91.4% | 93.5% | 60.2% | 91.7% | 97.3% | 69.9% | 98.8% |
| HI | 0.26 | 0.3 | 0.3 | 0.23 | 0.27 | 0.25 | 0.25 | 0.8 | 0.17 |
| Deudenum ($V0.5cc$) Gy | 27.96 | 29.37 | 28.6 | 1.35 | 0.35 | 0.33 | 28.94 | 22.53 | 30.67 |
| Stomach ($V0.5cc$) Gy | 14.7 | 15.1 | 9.9 | 0.16 | 0.23 | 0.29 | 0.03 | 0.06 | 0.15 |
| Esophagus ($V0.03cc$) Gy | 0.32 | 0.45 | 0.4 | 8.75 | 22.17 | 16.2 | 0.78 | 0.45 | 3.14 |
| Liver-GTV ($D_{700}$) Gy | 0.43 | 0.39 | 0.48 | 0.14 | 0.13 | 0.12 | 0.005 | 0.01 | 0.08 |

## 3.2 Evaluation of FLASH effect:

The FLASH effect of ADP-pRF was evaluated by comparing the FLASH doses of ADP pRF plans for dose thresholds $D_0$ of 1 Gy, referred to as FLASH-1Gy, to the initial IMPT plan designed on pCT. The FLASH -1Gy of case 1 is shown in Figure 4, while the results of case 2 and case 3 are presented in supplementary material. Figure 4 highlights the FLASH effect for case 1 across all scenarios. It shows the IMPT plan, the FLASH effect for the initial pRF plan on pCT, and the ADP pRF plan constructed on re-CT. The FLASH effect from the ADP pRF plan on re-CT demonstrates a dose distribution and dosimetrics closely matching the FLASH effect we observed from our previous work[34].

Setting $D_0$ to 1 Gy resulted in significant reductions in case 1, the plan showing the FLASH effect is termed FLASH-1Gy. The FLASH-1Gy treatment led to a decrease of 42.7% for $D_{max}$ of the stomach, 61% for $V_{21Gy}$ of liver-GTV, and a 35.5% reduction in $D_{mean}$ of PTV-GTV when compared to the IMPT on pCT. These reductions were largely consistent with the FLASH effect of the initial pRF plan on pCT, for instance, the $D_{mean}$ of PTV-GTV showed a comparable reduction of 35% in the initial pRF plan.

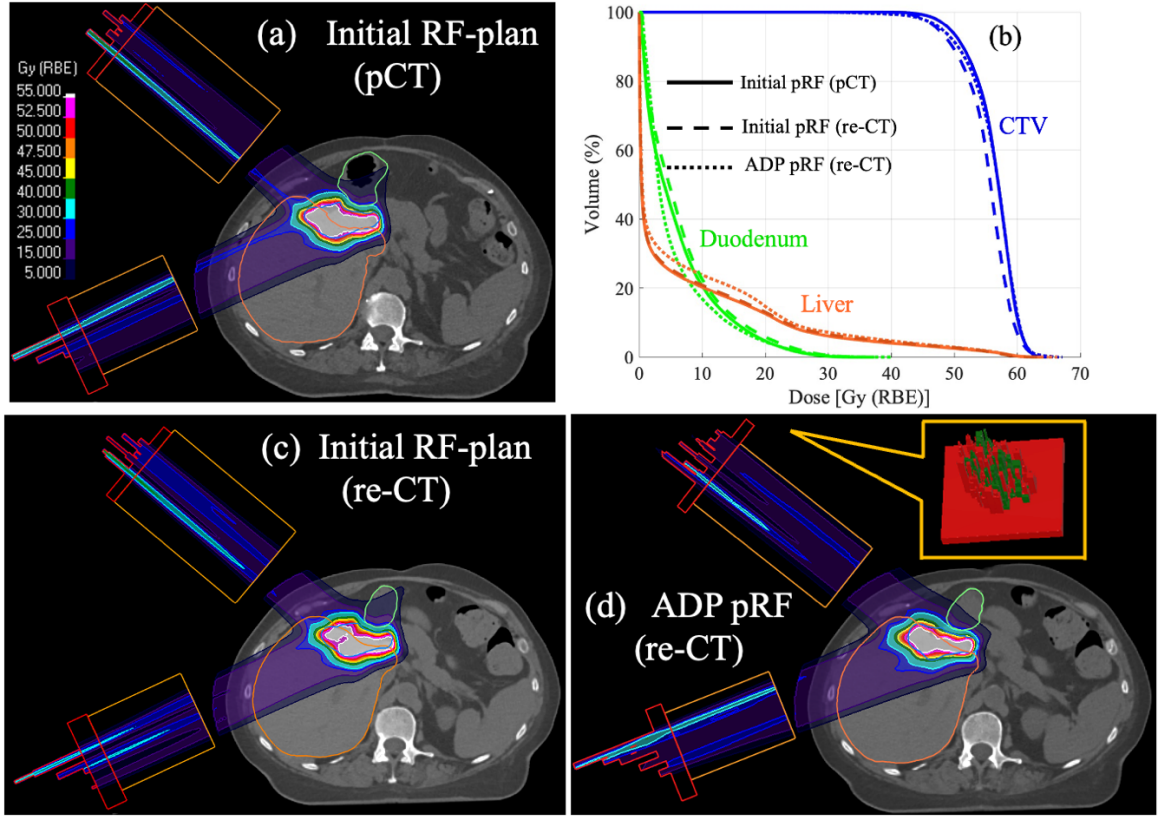

Figure 3. A two-beam single-energy proton plan utilizing designed Ridge Filters. (a) 2D dose distribution of the initial pRF plan on the original pCT, highlighting the target (CTV) outlined in blue and regions of interest (ROIs), with the liver outlined in orange and the duodenum in green. (b) Dose-volume histogram (DVH) comparison for the RF plans corresponding to the scenarios shown in (a), (c), and (d), presented as a percentage of the prescribed dose of 50 Gy. (c) 2D dose distribution of the initial pRF plan created with the Ridge Filter design based on the original CT. Anatomical and spatial changes in the CTV, liver, and duodenum during treatment lead to undercoverage of the target and increased radiation exposure to the duodenum. (d) 2D dose distribution of the ADP-pRF plan generated using the new adaptation method, demonstrated improved target coverage and dose conformity while effectively sparing the OARs.

Similarly, for case 2, $D_{mean}$ of PTV-GTV decreased by 33.1% in FLASH-1Gy compared to a 35.2% reduction when we observed the FLASH effect of the initial pRF plan. In case 3 a more pronounced FLASH effect was observed. The reduction in the $D_{mean}$ of PTV-GTV, and $D_{max}$ of deuodenum were 19.8 Gy and 8.8 Gy, respectively, compared to 17.8 Gy and 2 Gy observed in initial pRF.

At the dose threshold $D_0$ of 1 Gy, case 1 showed positive clinical benefits across most of the concerned normal tissues, with reductions in dose indicators varying from 35.5% to 61% when comparing FLASH-1Gy to conventional IMPT. Certain normal tissues displayed fewer therapeutic advantages in all three cases for FLASH-1Gy compared to initial pRF. For instance, in case 1, $V_{21Gy}$ of liver-GTV is 61% and 68.7% for FLASH-1Gy and initial pRF, respectively. Similarly, the $D_{max}$ of the duodenum for FLASH-1Gy is reduced by 40.3% compared to a 50.2% reduction in the initial pRF plan. However, in case 3 the dose indicators of OAR have shown better therapeutic advantages from the initial pRF plan, as the $D_{mean}$ of PTV-GTV and $D_{max}$ of duodenum showed a more pronounced FLASH effect as mentioned above.

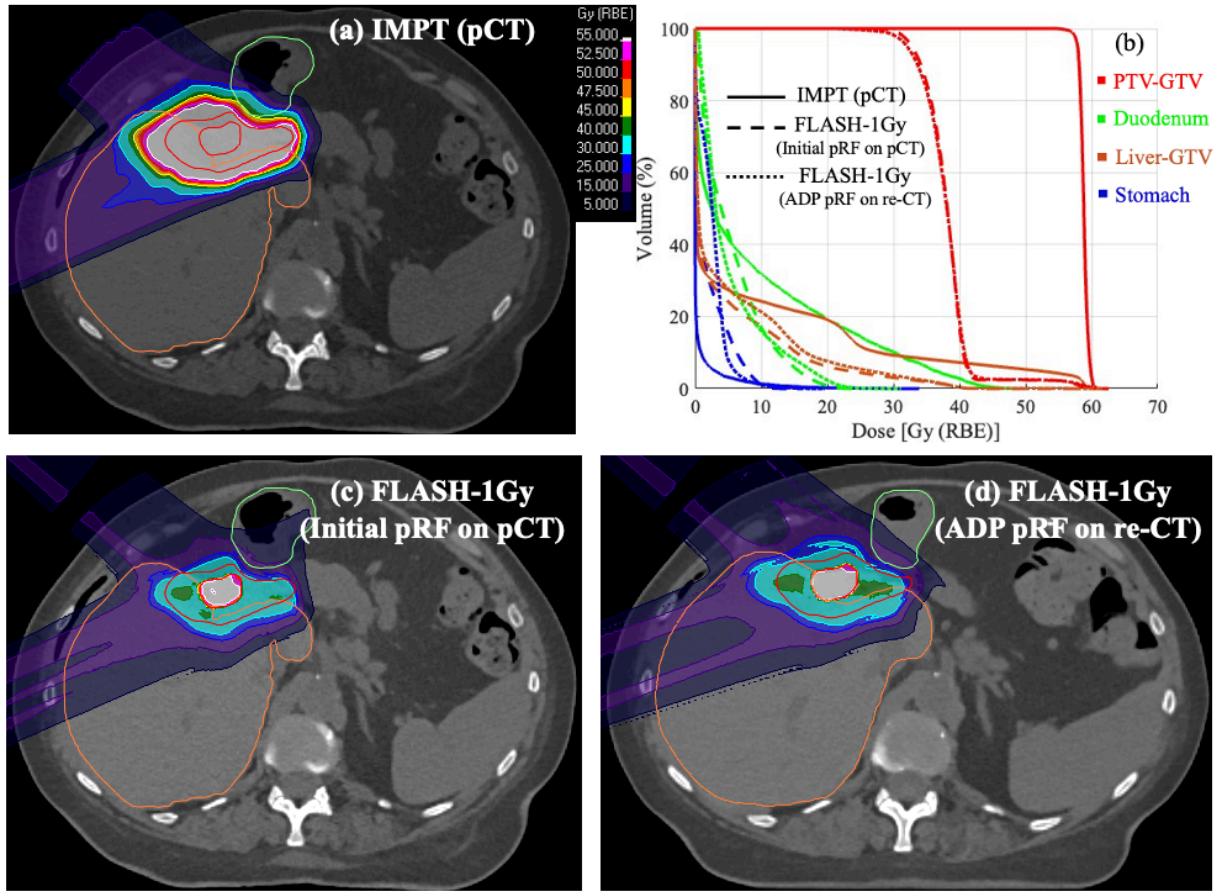

Figure 4. (a) 2D dose distribution of the IMPT plan on the planning CT (pCT). (b) Comparison of dose-volume histograms (DVHs) for IMPT on pCT (solid line), FLASH-1Gy on pCT (dashed line), and FLASH-1Gy on re-CT (dotted line), showing the percentage of the prescribed dose of 50 Gy delivered in 5 fractions. The mean liver dose and maximum dose to the duodenum (Dmax) are significantly lower in both FLASH plans than IMPT, as these critical organs are located further away in the re-CT. (c) FLASH-1Gy distribution from the initial pRF on the pCT, and (d) FLASH-1Gy distribution on the re-CT using the ADP pRF with the proposed method.

## 4. Discussion:

In this study, we validated our proposed method for recycling modularized pRFs in single-energy proton FLASH planning across three liver SBRT cases. The previously established technique of modularizing pRFs has simplified manufacturing processes and lowered costs by allowing the assembly of standardized libraries composed of fixed-size modules. Building on this foundation, our proposed framework extends the utility of these modules by recycling ridge pins, initially designed on pCT, even when anatomical changes necessitate adaptive replanning. This approach not only preserves the economic and logistical benefits of modularization but also addresses a critical challenge in adaptive FLASH therapy, i.e., the need for rapid, efficient recalibration of pRFs in response to anatomical variations.

The main reason for this study is a need to initiate an adaptation or recycling process arising from the application of initial pRFs to re-CT. Across all three cases, the V100 of the CTV decreased by a minimum of 4.5% (case 2), with the most severe reduction observed in case 3 (33.3%). The significant decline in dose coverage correlates with a significant dimensional difference between the CTV of pCT and re-CT. For instance, in case 3, the CTV in re-CT extended 2.13 cm superiorly compared to the CTV of pCT, resulting in the lowest dose coverage in the new target volume. These findings were further supported by HI values obtained from applying the initial pRF to re-CT. The HI values increased across all cases indicating an increase in dose heterogeneity, i.e., a larger variation in dose across the target volume. The largest increase in HI values was observed in case 3 by 0.55, reflecting substantially degraded dose homogeneity, attributed to greater dimensional variations in the CTV. These findings highlight that reapplying the initial pRFs without modification is impractical, as it would lead to clinically unacceptable plan quality. The traditional adaptive method would require a full 3D-printed redesign which is equally difficult due to time and cost constraints, emphasizing the need for a solution that encompasses adaptation as well as recycling on the initial pRF plan.

Our novel approach addressed these challenges by recycling pre-designed ridge pins from the initial pRF. By using the least square linear regression technique, we recycled 60–90% of the original pins. The proportion of recycled pins correlated inversely with the magnitude of CTV volume changes between scans. For example, in case 1, where the initial and new CTV volumes differed only marginally ($\Delta < 1\%$), the adaptive method recycled over 90% of the original pins. In contrast, case 3 exhibited a 35.3 cc increase in CTV volume, resulting in about 70% pin recovery. Case 2 demonstrated the most complex spatial changes: despite having the smallest initial CTV volume (21.35 cc), its repositioning in the re-CT caused both treatment beams to traverse increased lung volume. This anatomical shift necessitated the redesigning of 35% of the pins, though our method still retained 65% of the original pins. Nonetheless, our approach still allowed an average of 70% pin recycling across all these cases, underscoring the method's robustness to moderate anatomical variations.

After completing the ADP-pRF plan using recycled pins from the initial pRF plan, we conducted a dosimetric evaluation of its quality across all cases. In case 1, the ADP-pRF V100 for CTV was only 2.3% lower than the initial pRF plan, while case 2 showed a negligible 1.8% difference. Case 3 demonstrated a 1.5% improvement, suggesting potential robustness in scenarios with larger differences in the volume of CTV. All values remained within the BR001 protocol tolerance. HI values for ADP-pRF plans remained comparable to the initial plan indicating that the ADP plan can achieve similar dose uniformity as the initial pRF plan.

OAR doses in ADP-pRF plans were also generally comparable to initial pRF plans, minimizing unintended toxicity trade-offs. The Liver-GTV ($D_{700}$) increased slightly by 0.5 Gy and 0.8 Gy in cases 1, while remaining stable in case 2 and case 3. Notably, the Stomach (V0.5cc) improved significantly in case 1, decreasing by 4.8 Gy. The V0.03cc for the esophagus increased by 7.5 Gy in case 2 and 2.4 Gy in case 3, due to dose distribution extending closer to the esophageal region in re-CT of both cases.

Our iterative recycling process can also enable real-time plan quality assessment and parameter tuning during pin recycling by adjusting the objectives of ADP DS plans. The pRF resolution can also be customized via pin geometry and step count modifications. This flexibility enhances both manufacturing efficiency and dosimetric precision, offering a scalable solution for adaptive FLASH therapy.

The FLASH effect in monoenergetic pRF plans was examined using the FLASH effectiveness model. FLASH effectiveness model integrates the dose threshold, dose rate threshold, and FLASH persistence required to trigger the FLASH effect. The FLASH effect was quantified in all three cases by assessing the reduction in physical dose and the improvement in clinical outcomes, as demonstrated in our previous research[34]. Both initial pRF and ADP-pRF plans achieved comparable FLASH effect magnitudes, suggesting robustness in preserving clinical advantages during adaptation.

The use of a dose threshold ($D_0$ = 1 Gy) within our ADP FLASH framework resulted in a significant drop in normal tissue exposure while preserving GTV coverage. In Case 1, when comparing the FLASH effect induced by the initial pRF versus ADP-pRF, the PTV-GTV $D_{mean}$ demonstrates a reduction of 35% and 35.5%, respectively. This indicates a similar FLASH effect occurring in the tissues adjacent to the target area. $V_{21Gy}$ of liver-GTV recorded a 7.1% less reduction in the dose because the liver is in a low dose spillage region in re-CT which triggers less FLASH effect. Case 2 demonstrated similar consistency, with FLASH-1Gy achieving a 33.1% PTV-GTV Dmean reduction versus a 35.2% reduction in the initial pRF plan. Notably, case 3 revealed a better FLASH effect in the ADP-pRF plan, with absolute reductions in PTV-GTV Dmean (19.8 Gy vs. 17.8 Gy) and duodenum Dmax (8.8 Gy vs. 2 Gy) surpassing the initial pRF performance. Future work should explore adaptive $D_0$ adaptation tailored to anatomical and dosimetric shifts, ensuring a robust FLASH effect across diverse clinical scenarios.

# 5. Conclusion:

This study validated a method for recycling modularized pRFs in single-energy proton FLASH planning across SBRT cases. By using pre-designed ridge pins, our ADP pRF approach effectively retained up to 90% of the original pins while maintaining dosimetric integrity and reducing the need for a full 3D-printed redesign. The results demonstrated that ADP-pRF plans preserved CTV coverage, improved HI values, and minimized OAR toxicity trade-offs, with certain cases benefiting from enhanced FLASH effects. Despite minor variations in dose distributions, our findings underscore the potential to balance FLASH efficacy and OAR sparing. Future work should explore dose threshold ($D_0$) tuning to ensure robust and efficient FLASH therapy across diverse clinical scenarios. This framework offers a scalable and efficient solution for adaptive proton therapy, optimizing both clinical effectiveness and logistical feasibility.

**Conflicts of interest**

The authors confirm that there are no known conflicts of interest associated with this publication.

14   Newhauser, W. D. & Zhang, R. The physics of proton therapy. *Physics in Medicine & Biology* **60**, R155 (2015).

15   Jolly, S., Owen, H., Schippers, M. & Welsch, C. Technical challenges for FLASH proton therapy. *Physica Medica* **78**, 71-82 (2020).

16   Gao, H. *et al.* Simultaneous dose and dose rate optimization (SDDRO) of the FLASH effect for pencil-beam-scanning proton therapy. *Medical physics* **49**, 2014-2025 (2022).

17   Van De Water, S., Safai, S., Schippers, J. M., Weber, D. C. & Lomax, A. J. Towards FLASH proton therapy: the impact of treatment planning and machine characteristics on achievable dose rates. *Acta oncologica* **58**, 1463-1469 (2019).

18   van Marlen, P. *et al.* Bringing FLASH to the clinic: treatment planning considerations for ultrahigh dose-rate proton beams. *International Journal of Radiation Oncology\* Biology\* Physics* **106**, 621-629 (2020).

19   Folkerts, M. M. *et al.* A framework for defining FLASH dose rate for pencil beam scanning. *Medical physics* **47**, 6396-6404 (2020).

20   Lin, Y. *et al.* SDDRO-Joint: simultaneous dose and dose rate optimization with the joint use of transmission beams and Bragg peaks for FLASH proton therapy. *Physics in medicine & biology* **66**, 125011 (2021).

21   Ma, C. *et al.* Feasibility study of hybrid inverse planning with transmission beams and single-energy spread-out Bragg peaks for proton FLASH radiotherapy. *Medical Physics* **50**, 3687-3700 (2023).

22   Schwarz, M., Traneus, E., Safai, S., Kolano, A. & van de Water, S. Treatment planning for Flash radiotherapy: General aspects and applications to proton beams. *Medical Physics* **49**, 2861-2874 (2022).

23   Liu, R. *et al.* An integrated biological optimization framework for proton SBRT FLASH treatment planning allows dose, dose rate, and LET optimization using patient-specific ridge filters. *arXiv preprint arXiv:2207.08016* (2022).

24   Zhang, G., Gao, W. & Peng, H. Design of static and dynamic ridge filters for FLASH–IMPT: A simulation study. *Medical physics* **49**, 5387-5399 (2022).

25   Kang, M., Wei, S., Choi, J. I., Lin, H. & Simone II, C. B. A universal range shifter and range compensator can enable proton pencil beam scanning single-energy Bragg peak FLASH-RT treatment using current commercially available proton systems. *International Journal of Radiation Oncology\* Biology\* Physics* **113**, 203-213 (2022).

26   Ma, C. *et al.* Streamlined pin-ridge-filter design for single-energy proton FLASH planning. *Medical Physics* **51**, 2955-2966 (2024).
19